# Theoretical method for generating regular spatiotemporal pulsed-beam with controlled transverse-spatiotemporal dispersion

Zhaoyang Li* and Noriaki Miyanaga

Institute of Laser Engineering, Osaka University, 2-6 Yamada-oka, Suita, Osaka 565-0871, Japan.
*E-mail: zhaoyang-li@ile.osaka-u.ac.jp

## Abstract

Herein we theoretically report a method that generates a transverse-spatiotemporal dispersion (T-STD), which is distinct from previous spatial, temporal, and longitudinal-spatiotemporal optics dispersions. By modulating T-STD, two not yet reported spatiotemporally structured beams (STSBs), i.e., the honeycomb beam and the picket-fence beam, can be generated in the space-time domain. The generated STSBs have novel and tunable periodic distributions. T-STD, STSB and their inner physical relationship are analyzed and introduced. We believe that this method might open a path towards new optical beams for potential applications, such as ultrafast optical fabrication and detection.

## 1. Introduction

Optical dispersion is one of the fundamental phenomena in polychromatic, broadband and short-pulse light sources in optics [1]. Spatial dispersion was the first observed form of dispersion and is dependent on the wave vector. For example, after a single refraction or diffraction process, spatial dispersion can easily be generated and is used as a basic tool in fields such as spectroscopy [2]. Temporal dispersion is generally relevant to the frequency-dependent phase velocity in optical media or systems, and is commonly used to shape or modulate optical pulses in time to produce pulses, such as chirped pulses and solitons, for a wide range of applications [3, 4].

Recently, following the rapid developments in ultrashort optical pulse generation, some coupling effects were presented. Akturk et al. presented a general theory of first-order spatiotemporal coupling (STC) [5], i.e., coupling between the spatial (or spatial-frequency) and temporal (or frequency) coordinates of Gaussian pulses and beams. The electric field ($E$-field) of a first-order STC can be written in the form $E(x,t) \propto \exp\left(Q_{xx}x^2 + 2Q_{xt}xt - Q_{tt}t^2\right)$, and the cross-term $Q_{xt}$ provides information about eight types of STCs. In practical applications, wave-front rotation can be used in high-harmonic generation experiments to enable the production of isolated attosecond pulses [6]. Focusing of spatial dispersion beams with narrow initial beam apertures (i.e., frequency-dependent beams that are completely separated in the transverse direction), which is known as simultaneous space-time focusing, can be used to improve axial resolution for wide-field imaging applications [7-9]. Only very recently, with the help of STC, the velocities of ultrashort light pulses were controlled well [10], and a new focusing scheme of flying focus was demonstrated [11]. In general, first-order STCs and the related distortions show slow near-linear variations; however, cases involving higher-order STCs become more complex. For example, in the case of longitudinal-spatiotemporal dispersion involving second-order spatial and temporal terms in the nonlinear Schrödinger equation without the slowly-varying envelope approximation, the coefficients contain the spatial dispersion and group velocity dispersion information, which would then lead to wave envelopes with either relativistic or pseudo-relativistic characteristics [12].

## 2. Method and simulations

In this paper, we describe a rarely researched dispersion of T-STD, which is generated in an improved beam-shaping setup. Figure 1 shows a short pulse signal is injected into a 4-$f$ grating system, and the initial beam aperture is expanded to a size that is relatively equivalent to the spatial separation induced by the angular dispersion. A phase spatial light modulator (P-SLM) is positioned just in front of the first lens L1 instead of the Fourier plane FP, and a space-dependent phase modulation is introduced at this space-spectrum coupling plane. Before the first grating G1, each frequency has a planar phase-front, however, at the P-SLM, a space-dependent phase modulation is overlaid on each phase-front, although the frequency-dependent beams are separated along the transverse direction (x-axis). Then, after the second grating G2, where the angular dispersion is removed, the spectral phase (i.e., temporal dispersion) varies across the transverse direction, thereby T-STD is produced.

The differences in the proposed method when compared with three kinds of previous methods are clear. In the simultaneous space-time focusing, the beam aperture is usually very narrow when compared with the spatial separation induced by the angular dispersion [7-9]. In the short pulse shaping, the phase modulator is positioned at the Fourier plane (and not at the space-spectrum coupling plane) of the 4-*f* grating system, where the beam's spatial properties are completely removed. And, this is why this method is usually called Fourier transform pulse shaping [13]. And, in the beam smoothing for high energy and power lasers, the transverse phase modulation is applied to a large-aperture beam, however there is usually zero or negligible spatial/angular dispersion [14, 15]. Consequently, the results in this paper were not found in previous simulations or experiments.

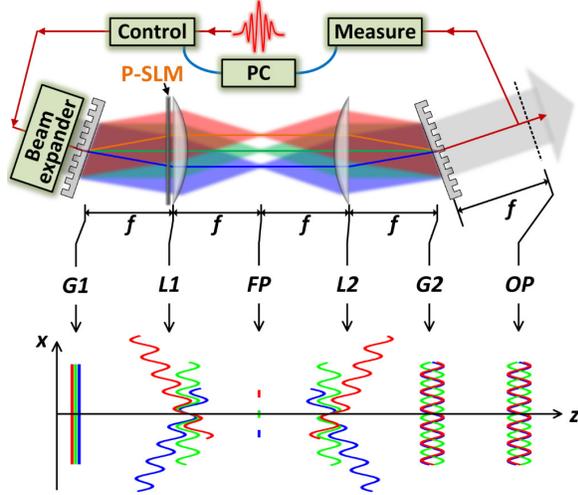

Fig. 1. A phase spatial light modulator (P-SLM) is inserted in a 4-f grating system to generated T-STD, and closed-loop spectral phase control and measurement are used to control the initial temporal dispersion. Lower schematically illustrates the phase-fronts of three frequencies at different positions in the setup. G1: grating 1, L1: lens 1, FP: Fourier plane, L2: lens 2, G2: grating 2, and OP: output plane.

We could simulate the complex amplitude of each frequency along the propagation in the setup by using the Collins diffraction formula

$$U_2(x_2, z, \omega) = \frac{\exp(ikz)}{i\lambda B} \int U_1(x_1, \omega) \cdot \exp\left[\frac{ik}{2B}(Ax_1^2 - 2x_1 x_2 + Dx_2^2)\right]dx_1 \quad (1)$$

where *A*, *B* and *D* are elements of the ABCD propagation matrix.

The simulation in this paper is based on the following parameters: Gaussian spectrum/pulse with a center wavelength of 1030 nm and a bandwidth of 20 nm (Full width at half maximum, FWHM), super-Gaussian beam with an order of 10 and a diameter of ~3 mm (Full width, FW), grating density of 200 g/mm, Littrow incident angle, focal length of 100 mm, and sine function phase modulation (from P-SLM) with initial modulation PV of $0.9\pi$ and period of 0.6 mm.

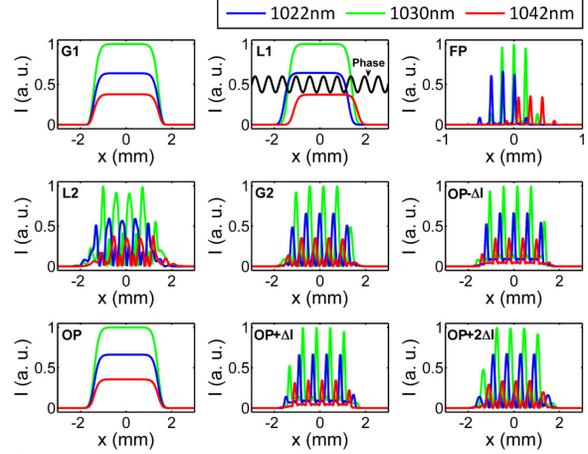

Fig. 2. Evolution of intensity spatial profiles of three frequencies (1022, 1030 and 1042 nm) at different positions in the setup. G1: grating 1, L1: lens 1, FP: Fourier plane, L2: lens 2, G2: grating 2, OP: output plane, and $\Delta l$ is the offset length (50 mm here).

Along the propagation path in the setup, Fig. 2 shows the intensity spatial profiles of three frequencies (1022, 1030 and 1042 nm) at several key positions in the setup. At the first grating G1, frequency-dependent beams have a same spatial location, however, at the first lens L1, they are separated along the angular dispersion direction (x-axis). In this case, the space-dependent phase modulation (from P-SLM) would modulate the phase-fronts of different frequencies differently. Because of the phase-front modulation and the propagation diffraction, the intensity profiles at the Fourier plane FP, the second lens L2 and the second grating G2 are distorted. However, at the output plane OP, which is the image position of the P-SLM, the spatial distortion of the intensity profile disappears. Consequently, at this position the amplitude/intensity distortion induced by the phase-front modulation and the propagation diffraction becomes negligible. However, for other offset positions away from the output plane OP, this condition cannot be satisfied. Meanwhile, for the ideal case, as a benefit of the use of the 4-*f* system, if the P-SLM does not work, the spatial and temporal dispersions induced by G1 could be compensated completely by G2, and the output is then restored to match the input without additional spatial, temporal or spectral modulations. In this case, the generated T-STD can be controlled well using the P-SLM. The amplitude modulation at the output plane OP can be neglected, and the pulsed-beam shaping is mainly determined by the phase modulation, i.e., T-STD.

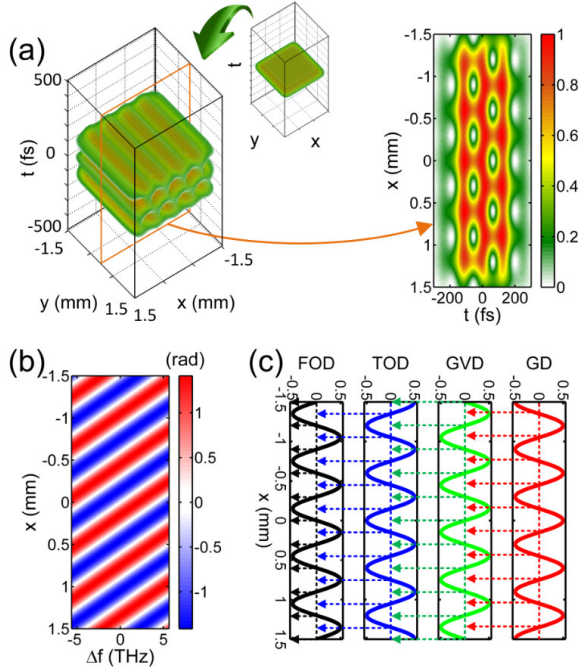
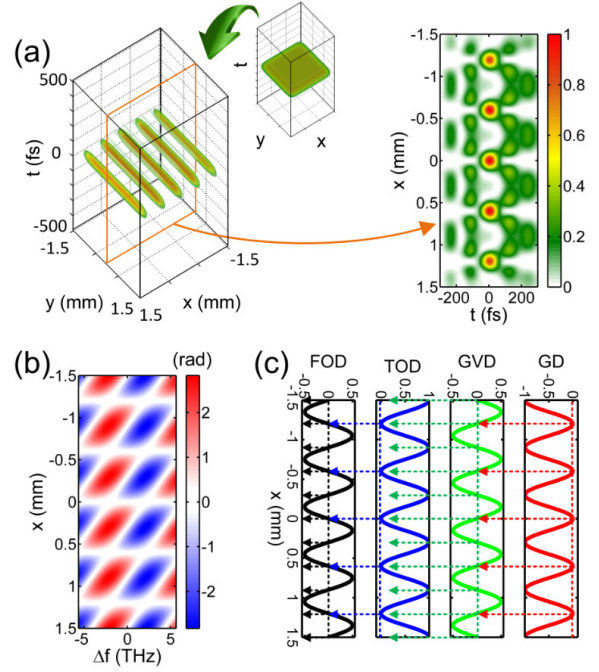

Fig. 3. Without the initial temporal dispersion control, (a) normalized intensity, (b) space-dependent spectral phase and (c) normalized spatiotemporal dispersion distributions in the space-time or space-spectrum domain. In (c), the zero-GD/TOD (odd orders) and the zero-GVD/FOD (even orders) are located at different spatial positions with staggered periodic locations, and, in (a), the input pulsed beam (super-Gaussian in space and Gaussian in time) is changed into a honeycomb beam. For observation, only $I>0.3$ are plotted in the 3D figure in (a).

Using the above parameters, we simulate the propagation of each frequency by the Collins diffraction formula, and, after the coherent addition, the pulsed beam in both time and space is obtained by the Fourier transform at each spatial position. Figure 3(a) shows the input pulsed beam (super-Gaussian in space and Gaussian in time) is changed into a honeycomb beam in the space-time domain, and the spatiotemporal distribution is in the x-t plane. Fig. 3(b) shows the sine function phase modulation in space (x-axis) is tilted along the spectral frequency axis (Δf-axis).

Fig. 4. With the initial temporal dispersion control, (a) normalized intensity, (b) space-dependent spectral phase and (c) normalized spatiotemporal dispersion distributions in the space-time or space-spectrum domain. In (c), the zero-GD/TOD (odd orders) and the zero-GVD/FOD (even orders) have same locations at $x=0, \pm 0.6, \pm 1.2$ mm, and, in (a), the input pulsed beam (super-Gaussian in space and Gaussian in time) is changed into a picket-fence beam. For observation, only $I>0.3$ are plotted in the 3D figure in (a).

Figure 1 shows if an appropriate space-independent initial temporal dispersion (i.e., spectral phase) is loaded, for example the spectral phase of the center optical ray is measured and pre-compensated using commercial products Wizzler and Dazzler by Fastlite Inc. [16], the dispersion-free condition would occur at the beam center as well as some positions with certain periodic locations along the transverse direction (x-axis) [Fig. 4(b)], and then the Fourier transform-limited (FTL) pulse would be achieved at these positions. However, at other positions, temporal dispersions and pulse distortions cannot be avoided. In this case, Fig. 4(a) shows the input pulsed beam is then changed into a picket-fence beam in the space-time domain, and the spatiotemporal distribution is also in the x-t plane. Now, two spatiotemporally structured beams, i.e., a honeycomb beam and a picket-fence beam, are generated theoretically with the proposed method.

## 3. Analytical expression of T-STD

Actually, the physical mechanism of the generation of STSB can also be explained by T-STD. Two coordinate systems of x-z and x′-z′ are set up at the planes in front of the first grating G1 and at the P-SLM (near L1), $x$ and $x'$ are transverse directions, and $z$ and $z'$ are propagation directions of the central ray of the center frequency. The geometrical relationship between $x$ and $x'$ for a ray of arbitrary frequency within the beam is given by

$$x' = x\frac{\cos\theta_0}{\cos\alpha} - \left(f + x\frac{\sin\theta_0}{\cos\alpha}\right)\Delta\theta, \quad (2)$$

where $\alpha$ is the angle of incidence, $\theta_0$ is the diffraction angle of the center frequency $\omega_0$, $\Delta\theta$ is the diffraction angle deviation of an arbitrary frequency $\omega$ with respect to the center frequency $\omega_0$, and $f$ is the focal length. By calculating the derivatives of the first-order grating equation $\sin\alpha + \sin\theta_0 = 2\pi c/(\omega d)$ with respect to $\omega_0$, $\Delta\theta$ is given by $-2\pi c/(\omega_0^2 d\cos\theta_0)(\omega-\omega_0)$, where $d$ is the grating constant. The phase modulation function of the P-SLM is space-dependent $\phi(x') = h/2 \cdot \sin(2\pi x'/l)$, where $h$ and $l$ are the modulation PV and period, respectively. By substitution of Eq. (2) into $\phi(x')$, the spatio-spectral-phase modulation $\phi(x,\omega)$ is given by

$$\phi(x,\omega) = \frac{h}{2}\sin\left[Mx + N\cdot(\omega-\omega_0)\right], \quad (3)$$

where $M$ and $N$ satisfy

$$M = \frac{2\pi\cos\theta_0}{l\cos\alpha}, \quad (4)$$

and

$$N = \left(f + x\frac{\sin\theta_0}{\cos\alpha}\right)\frac{4\pi^2 c}{l\omega_0^2 d\cos\theta_0}. \quad (5)$$

Based on the angle sum identity and the Taylor expansion of the trigonometric functions, Eq. (3) can be rewritten as a Taylor expansion about $\omega_0$, as follows:

$$\phi(x,\omega) = \sum_{k=0}^{\infty}\left[\begin{array}{c}\frac{\phi^{(2k)}(x)}{(2k)!}(\omega-\omega_0)^{2k} \\ +\frac{\phi^{(2k+1)}(x)}{(2k+1)!}(\omega-\omega_0)^{2k+1}\end{array}\right], \quad (6)$$

where the coefficients $\phi^{(2k)}(x)$ and $\phi^{(2k+1)}(x)$ satisfy

$$\phi^{(2k)}(x) = \frac{h}{2}\sin(Mx)(-1)^k N^{2k}, \quad (7)$$

and

$$\phi^{(2k+1)}(x) = \frac{h}{2}\cos(Mx)(-1)^k N^{2k+1}. \quad (8)$$

respectively. Equations (7) and (8) show the even and odd orders of the T-STD, e.g., space-dependent group delay (GD), group velocity dispersion (GVD), third-order dispersion (TOD), fourth-order dispersion (FOD), etc. When an appropriate space-independent initial temporal dispersion is considered, correction terms for $-\phi_{x0}^{(2k)}$ and $-\phi_{x0}^{(2k+1)}$ should be added to the even and odd orders of the T-STD in Eqs. (7) and (8), respectively. $x_0$ is actually one of the dispersion-free positions that can be an arbitrary spatial position in the beam transverse direction. For example, in the closed-loop spectral phase control and measurement process, because both of controller and detector (e.g., Wizzler and Dazzler from Fastlite Inc.) have no spatial resolution, and $x_0$ is the exact position of the optical ray is injected into the detector.

Then, the phase modulation can be written as $\exp[i\phi(x,\omega) - i\phi_{x0}(\omega)]$, which consists of the T-STD $\phi(x,\omega)$ and the initial temporal dispersion $-\phi_{x0}(\omega)$. We should emphasize here that $\phi(x,\omega)$ is a 2D data matrix (where rows and columns denote space and spectral frequency, respectively) obtained using Eq. (3), and $-\phi_{x0}(\omega)$ is also a 2D data matrix in which each row is equal to $\phi(x_0,\omega)$, i.e., the data in the matrix is space-independent. Using Eqs. (7) and (8), the $x$-dependent GD, GVD, TOD, and FOD can be directly calculated, and Fig. 3(c) shows that, without the initial temporal dispersion, the zero-GD and the zero-TOD (the odd orders of the T-STD) are located at the same spatial positions of $x = 1/2(k+1/2)l$, $k = 0, \pm 1, \pm 2 \cdots$ [from Eq. (8)]; however, the zero-GVD and the zero-FOD (the even orders of the T-STD) shows different locations of $x = 1/2kl$, $k = 0, \pm 1, \pm 2 \cdots$ [from Eq. (7)]. But, the locations of the zero-odd and zero-even orders of the T-STD have staggered periodic distributions, and consequently a periodically structured light (i.e., honeycomb beam) appears in space and time. When the initial temporal dispersion detected at $x_0 = 0$ is loaded, Fig. 4(c) shows the zero-GD, the zero-GVD, the zero–TOD, and the zero-FOD (all orders of the T-STD) have the same locations of $x = kl$, $k = 0, \pm 1, \pm 2 \cdots$, where the dispersion-free condition, i.e., the FTL pulse, can be obtained. Thereby, the input pulsed beam (super-Gaussian in space and Gaussian in time) is changed into a picket-fence beam.

## 4. Controllability of STSB

According to our simulation, the controllability of the honeycomb beam is limited, however that of the picket-fence beam is very high. Equation (4) shows that the period of the trigonometric functions in both Eqs. (7) and (8) is $l\cos\alpha/\cos\theta_0$, and in this case, the period $w_2$ of the picket-fence beam can be adjusted conveniently by directly changing the modulation period $l$. Because the Littrow incident angle is used, the red dot-line in Fig. 5(b) shows the picket-fence period $w_2$ equals the modulation period $l$. Figure 5(a) shows the 3D intensity profile of the picket-fence beam presented in Fig. 4(a) in the $x$-$t$ domain. We define the duty cycle as the ratio of the picket-fence width $w_1$ (FWHM) to the picket-fence period $w_2$ at $t=0$. Equations (5), (7) and (8) show that the even and odd orders of the T-STD, $\phi^{(2k)}(x)$ and $\phi^{(2k+1)}(x)$, are proportional to $l^{-2k}$ and $l^{-(2k+1)}$, respectively, and then the T-STD, accordingly the duty cycle of the picket-fence beam, can be increased by reducing the modulation period $l$.

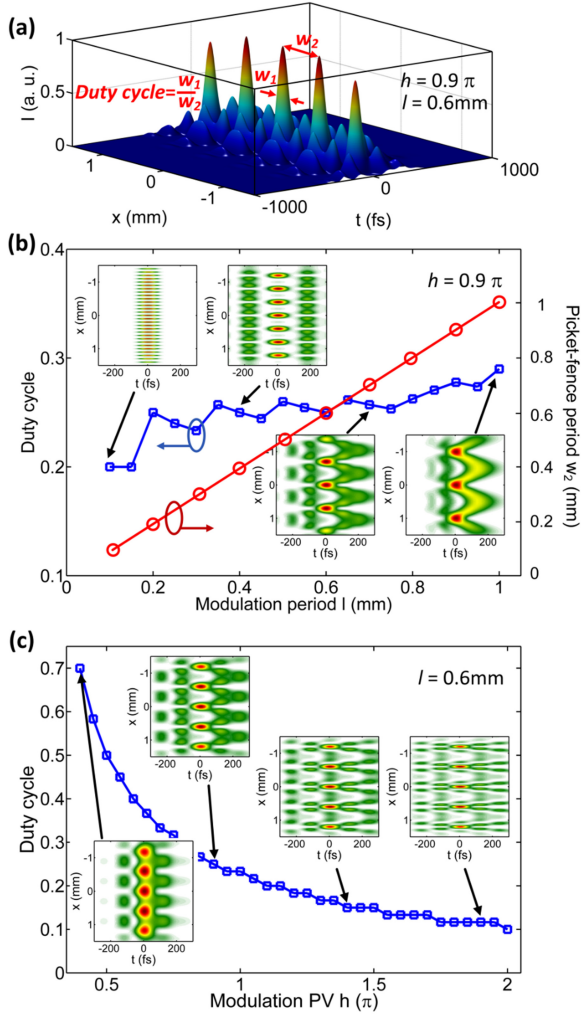

Fig. 5. (a) Intensity distribution of the picket-fence beam in the $x$-$t$ domain for $h=0.9\pi$ and $l=0.6$ mm. The duty cycle is defined as the ratio of the picket-fence width (FWHM) to the picket-fence period at $t=0$. The duty cycle and the spatial period of the picket-fence beam are shown as functions of the modulation (b) period $l$ and (c) PV $h$. Insets in (b) and (c) show the intensity spatiotemporal distributions corresponding to different modulation parameters.

Equations (7) and (8) show that the T-STD is also proportional to $h$, consequently the duty cycle can be adjusted by varying the modulation PV $h$, too. Figures 5(b) and 5(c) illustrate the evolution of the duty cycle with variation of the modulation period $l$ and PV $h$. When the modulation period increases from 0.1 to 1 mm, the duty cycle gradually increases from around 0.2 to 0.3, and some oscillations can be observed. However, by increasing the modulation PV from $0.4\pi$ to $2\pi$, the duty cycle can be reduced dramatically from around 0.7 to 0.1 in an almost linear manner. In this case, when compared with the modulation period $l$, the modulation PV $h$ is preferred for use in duty cycle adjustments.

Consequently, we can conclude that the period $w_2$ and the duty cycle of the picket-fence beam can be controlled conveniently by simply adjusting the modulation period $l$ and PV $h$, respectively. Additionally, insets in Figs. 5(b) and 5(c) are corresponding intensity spatiotemporal distributions for different modulation period $l$ and PV $h$, which shows a clean picket-fence beam could be generated when a short modulation period $l$ and a high modulation PV $h$ are used.

## 5. Conclusion

In conclusion, we have theoretically presented a rarely researched dispersion in the form of the T-STD along with the generation of not yet reported STSBs. By controlling T-STD, STSBs, i.e., the honeycomb beam and the picket-fence beam, can be produced in the space-time domain. For the honeycomb beam, to produce a uniform distribution, a careful optimization of the modulation function (provided at the space-spectrum coupling plane) is required. However, for the picket-fence beam, its period and duty cycle can be controlled well via the modulation function. It should be noted that this T-STD method provides new potentials for coherent control in optics, and it also produces new types of STSBs with high and practical tenability for potential applications.